
\magnification=\magstep1
\hoffset=0.1truecm
\voffset=0.1truecm
\vsize=23.0truecm
\hsize=16.25truecm
\parskip=0.2truecm
\def\newpage{\vfill\eject}

\def\fun#1#2{\lower3.6pt\vbox{\baselineskip0pt\lineskip.9pt
  \ialign{$\mathsurround=0pt#1\hfil##\hfil$\crcr#2\crcr\sim\crcr}}}

\def\vol{ {L^3} }
\def\ratA{ {\cal R}_4 }
\def\ratB{ {\cal R}_3 }
\def\seA{ S_{E4} }
\def\seB{ S_{E3} }
\def\eff{ {\varepsilon} }
%
%
$\,$
\vskip 0.50truein
\centerline{\bf GENERAL SOLUTIONS FOR TUNNELING}
\vskip 0.05truein
\centerline{\bf OF SCALAR FIELDS WITH QUARTIC POTENTIALS}
\vskip 0.2truein
\centerline{\bf Fred C.~Adams}
\vskip 0.1truein
\centerline{\it Physics Department, University of Michigan}
\centerline{\it Ann Arbor, MI  48109}

\vskip 0.4truein
\centerline{submitted to {\it Physical Review} {\bf D\/} }
\vskip 0.4truein
\centerline{\it 9 February, 1993}
\vskip 0.4truein

\centerline{\bf ABSTRACT}
\vskip 0.2truein

For the theory of a single scalar field $\varphi$ with a quartic
potential $V(\varphi)$, we find semi-analytic expressions for the
Euclidean action in both four and three dimensions.  The action
in four dimensions determines the quantum tunneling rate at zero
temperature from a false vacuum state to the true vacuum state;
similarly, the action in three dimensions determines the thermal
tunneling rate for a finite temperature theory.
We show that for all quartic potentials, the action can be
obtained from a one parameter family of instanton solutions
corresponding to a one parameter family of differential equations.
We find the solutions numerically and use polynomial fitting
formulae to obtain expressions for the Euclidean action.  These
results allow one to calculate tunneling rates for the entire
possible range of quartic potentials, from the thin-wall (nearly
degenerate) limit to the opposite limit of vanishing barrier height.
We also present a similar calculation for potentials containing
$\varphi^4 \ln \varphi^2$ terms, which arise in the one-loop
approximation to the effective potential in electroweak theory.


\newpage
\centerline{\bf I. INTRODUCTION}
\medskip

The early universe may have experienced a series of first order
phase transitions.  In these phase transitions, scalar fields
can be an important component of the universe and thus the
tunneling of scalar fields from one vacuum state to another
plays an important role. Once a field configuration becomes
trapped in a metastable state (the false vacuum), bubbles of
the true vacuum state nucleate in the sea of false vacuum and
begin growing spherically. The basic problem is to calculate
the tunneling rate (the decay probability) from the false vacuum
state to the true vacuum state, i.e., the bubble nucleation rate
per unit time per unit volume.
For tunneling of scalar fields at zero temperature (generally
called quantum tunneling), the four-dimensional Euclidean action
of the theory largely determines this tunneling rate.  For
tunneling at finite temperature (thermal tunneling), the three
dimensional Euclidean action must be determined. In this paper,
we derive semi-analytic expressions for the Euclidean action in
both four and three dimensions for the general case of a single
scalar field with a quartic potential.

In the original version of the inflationary universe [1],
the phase transition which ends the inflationary epoch
is first order and proceeds through the quantum tunneling
of scalar fields from a false vacuum state to a true vacuum
state.  Although this version of inflation has been shown to
have problems [2], many alternate models which also occur through
first order phase transitions have been proposed.  These models
include extended inflation [3], double field inflation [4],
hyperextended inflation [5], and many others.  The exact
tunneling rates are important for inflationary models
because these rates determine whether or not the phase
transition can complete [2] and also determine the size spectrum
of bubbles (and hence cosmological perturbations) produced [6].

In addition to the inflationary universe paradigm, other first
order phase transitions might occur in the early universe.
For example, many papers have explored possible consequences
of a first order electroweak phase transition, in particular
the possibility of generating the observed baryon asymmetry
of the universe [7].  The possibility that the electroweak
vacuum is unstable has also been proposed and the tunneling
rates have been used to constrain particle masses in the
Weinberg-Salam model [8].

Although tunneling of scalar fields can be important,
relatively little work has been done on calculating general
tunneling rates.  Previous work [9, 10, 11] has shown
that this tunneling rate can be calculated using semiclassical
techniques (these are reviewed in \S II).
The original calculation [9, 10] focused on the case of
nearly degenerate vacua (i.e., a small energy difference
between the two vacuum states) and found a closed form analytic
solution for the tunneling rate.  The limit of nearly degenerate
vacua is also called the ``thin wall limit'' because, in this limit,
the length scale over which the field $\varphi$ changes its value from
one vacuum to another is short compared to the radius of the bubble.
Unfortunately, except for this solution in the thin wall limit,
closed form solutions for the Euclidean action are rare.
Many authors simply use the solution for the thin wall limit,
even though the vacua in the problem are not sufficiently
equal in energy for the approximation to be valid.  The other
alternative is to calculate numerically the Euclidean action and
hence the tunneling rate for specific potentials [8, 12].
The case of tunneling without potential barriers (i.e., with
perfectly flat potentials) has also been considered [13].
Recently, the particular cases of triangle and square potentials
have been calculated analytically [14].
All of the results discussed thus far apply to tunneling at
zero temperature, i.e., quantum tunneling.  The theory of
tunneling has been generalized to include the decay of the
false vacuum at finite temperatures [15]. As before, closed
form solutions for the Euclidean action are generally not
available.

In this paper, we provide a more general solution to this problem.
We first show that for a general class of quartic potentials with
two minima, the problem can be reduced to a one parameter family;
the remaining parameters in the potential are scaled
out of the problem.  We then find the solutions for this entire one
parameter family using numerical techniques and derive polynomial
fitting formulae to express the results. We thus obtain semi-analytic
expressions for the Euclidean action in both four and three dimensions;
these expressions can be used to determine the tunneling rate
for this entire class of quartic potentials with two minima.
We also perform this same calculation for quartic potentials
containing logarithmic terms, such as those which arise in
the effective potential of electroweak theory.

This paper is organized as follows.  In \S II, we review the
semiclassical formalism used to calculate the tunneling rates
[9, 11, 15].  In \S III, we study the case of a general quartic
potential with two minima and find semi-analytic expressions for
the Euclidean action in both three and four dimensions; these
expressions are valid for the entire class of potentials.
In \S IV, we consider the case of quartic potentials with
logarithmic terms, such as those which arise in electroweak
theory. We conclude in \S V with a discussion of our results.

\bigskip
\centerline{\bf II. BASIC FORMULATION}
\medskip

In this section, we quickly review the formalism used for
calculating tunneling probabilities [9, 10, 11, 15].  This
discussion applies to the theory of a single scalar field
$\varphi$ with a potential $V(\varphi)$ which has both a false
vacuum state (a metastable state) and a true vacuum state.
The quantity of physical interest is the decay probability
$\Gamma/ \vol$ per unit time per unit volume.  We consider
both the cases of tunneling at zero temperature and at
finite temperature.

\bigskip
\centerline{\bf A. Quantum Tunneling at Zero Temperature}
\medskip

As shown by
previous authors [9], in the zero temperature limit this
quantity can be written (to leading order in $\hbar$) in the form
$${\Gamma \over \vol} = K {\rm e}^{-\seA} \, , \eqno(2.1)$$
where $K$ is a determinental factor [16] and where $\seA$ is
the Euclidean action in four dimensions (see Eq. [2.6] below).
Since the factor $K$ is generally of order $\eta^4$ (where $\eta$
is the energy scale of the phase transition -- see Refs. [9--11])
and the nucleation rate depends exponentially on the action $\seA$,
we will focus on the calculation of the action in this paper.

In order to compute the action $\seA$, we must find the solutions
to the Euclidean equations of motion for the field
$$\Bigl\{ {\partial^2 \over \partial t_E^2} + \nabla^2 \Bigr\}
\,  \varphi = {\partial V \over \partial \varphi} \, , \eqno(2.2)$$
where $t_E$ denotes the Euclidean time coordinate [17].
Although, in principle, all possible solutions of Eq. (2.2)
contribute to the tunneling process, Coleman [9] has shown
that the solution with the least action dominates.  Furthermore,
the solution with the least action has $O(4)$ symmetry [18].
As a result, we must solve the following $O(4)$ symmetric Euclidean
equation, which takes the deceptively simple form
$${d^2 \varphi \over dr^2 } + {3 \over r} {d \varphi \over dr}
= {\partial V \over \partial \varphi} , \eqno(2.3)$$
where $r$ is the four dimensional radial coordinate.

The appropriate boundary conditions for this problem [9] are
$${d \varphi \over dr} = 0 \qquad {\rm at} \quad r = 0 , \eqno(2.4)$$
and $$\varphi = \varphi_F
\qquad {\rm as} \quad r \to \infty \, , \eqno(2.5)$$
where $\varphi_F$ denotes the value of the field in the false vacuum
state. The boundary condition (2.4) is required to keep the action
nonsingular at the origin of coordinates $r=0$.  The condition
$\varphi = \varphi_F$ as $r \to \infty$ is really a combination of two
conditions: The first is that we want the instanton solution
(also called the bounce) to go from the false vacuum state
($\varphi = \varphi_F$) as $t_E \to - \infty$ and then back to the false
vacuum state as $t_E \to + \infty$. The second condition is
that we must also require $\varphi \to \varphi_F$ at spatial infinity
($| {\bf x} | \to \infty$) in order to keep the action finite.
These two conditions become Eq. (2.5) for the $O(4)$ symmetric
theory (see Ref. [9] for further discussion).

The desired Euclidean action then takes the form
$$\seA = 2 \pi^2 \int _0^\infty \, r^3 \, dr \, \Bigl[
{1 \over 2} \Bigl( {d \varphi \over dr} \Bigr)^2 + V[\varphi(r)]
\Bigr] \, , \eqno(2.6)$$
where $\varphi(r)$ is the $O(4)$ symmetric solution discussed
above.

\bigskip
\centerline{\bf B. Thermal Tunneling at Finite Temperature}
\medskip

The expression (2.1) is only valid at zero temperature.  At finite
temperature, the analogous expression for the decay probability
per unit time per unit volume takes the form
$${\Gamma_T \over L^3} = K_T \, {\rm e}^{-\seB/T} \,  , \eqno(2.7)$$
where $T$ is the temperature, $K_T$ is a new temperature dependent
determinental factor and $\seB$ is the Euclidean action in three
dimensions [15].  Typically we expect $K_T = {\cal O}(T^4)$
and we again focus only on the calculation of the Euclidean action
in this paper.

To evaluate the action $\seB$, we must solve the
$O(3)$ symmetric equation of motion
$${d^2 \varphi \over dr^2 } + {2 \over r} {d \varphi \over dr}
= {\partial V(\varphi, T) \over \partial \varphi} , \eqno(2.8)$$
where $r$ is now the usual radial coordinate in three dimensions.
Keep in mind that the potential $V(\varphi, T)$ appearing in equation (2.8)
generally contains temperature dependent terms and is not the same
as the zero temperature potential in the previous subsection.
Various expressions for the potentials in finite temperature
field theory are discussed in Refs. [19].

The boundary conditions for this problem are
$${d \varphi \over dr} = 0 \qquad {\rm at} \quad r = 0 , \eqno(2.9)$$
and
$$\varphi = \varphi_F \qquad {\rm as} \quad r \to \infty \, , \eqno(2.10)$$
where $\varphi_F$ denotes the location of the false vacuum.
Notice that these boundary conditions are almost the same as those
of Eqs. (2.4) and (2.5); the only difference is that $r$ is now the
radial coordinate in three dimensions.
The corresponding three dimensional action is
$$\seB = 4 \pi \int _0^\infty \, r^2 \, dr \, \Bigl[
{1 \over 2} \Bigl( {d \varphi \over dr} \Bigr)^2 +
V [\varphi(r), T] \Bigr] \, , \eqno(2.11)$$
where $\varphi(r)$ is the $O(3)$ symmetric
solution to equation (2.8).

\bigskip
\bigskip
\bigskip
\centerline{\bf III. GENERAL SOLUTIONS FOR QUARTIC POTENTIALS}
\medskip

In this section we obtain semi-analytic expressions for
the Euclidean action for a general class of quartic
potentials with two minima.  We begin by considering the
most general form for such a potential,
$$V(\varphi) = \lambda \varphi^4 - a \varphi^3 + b \varphi^2 +
c \varphi + d \, , \eqno(3.1)$$
where we have explicitly written the coefficient of the cubic term
with a minus sign (here, $a > 0$) in order to make the location
of the true vacuum occur for positive $\varphi$ [20].
We are interested in potentials having two minima
which correspond to a false vacuum and a true vacuum.
Without loss of generality, we can place the location
of the false vacuum at the origin ($\varphi = 0, V = 0$);
we can therefore set $d = 0 = c$.
The choice $d = 0$ is allowed because the constant term
in the potential does not enter into the dynamics [21]; the
choice $c = 0$ is allowed because we can make a translation
of the field $\varphi \to \varphi + \eta$ to place the minimum of
the potential at $\varphi = 0$. Notice that we must have
$\lambda > 0$ to keep the potential bounded from below and
we must have $b > 0$ to make the potential a minimum
(rather than a maximum) at $\varphi$ = 0.

Thus far, the equation of motion we need to solve takes the form
$${d^2 \varphi \over dr^2 } + {N-1 \over r} {d \varphi \over dr}
= 4 \lambda \varphi^3 - 3 a \varphi^2 + 2 b \varphi \, , \eqno(3.2)$$
where $r$ is the radial coordinate in $N$ dimensions.
In this paper, we consider both the case in which $N=4$
for the $O(4)$ symmetric theory and the case in which
$N=3$ for the $O(3)$ symmetric theory.
The equation of motion thus has three unspecified constants
($\lambda$, $a$, and $b$).  However, we can rescale
both the radial coordinate $r$ and the field $\varphi$.
If we adopt the judicious choice of new variables
$\phi$ and $\xi$ defined according to
$$\phi \equiv { 4 \lambda \over a } \varphi \qquad {\rm and}
\qquad \xi \equiv {a \over 2 \lambda^{1/2} } \, r \, , \eqno(3.3)$$
the differential equation can be put in a ``standard form''
$${d^2 \phi \over d \xi^2 } + {N-1 \over \xi} {d \phi \over d\xi}
= \phi^3 - 3 \phi^2 + \delta \phi , \eqno(3.4)$$
where the remaining parameter $\delta$ is defined by
$$\delta \equiv 8 \lambda b / a^2 \, . \eqno(3.5)$$
Notice that $b$ has units of $(energy)^2$ and $a$ has
units of $(energy)$ so that the parameter $\delta$ is
dimensionless.  Notice also that the choice of numerical
coefficients in the transformation of equation (3.3) is
arbitrary; different choices would lead to different
numerical coefficients on the right hand side of
Eq. (3.4).  The boundary conditions now can be written
$${d\phi \over d\xi} = 0 \qquad {\rm at} \quad \xi = 0 \, ,
\eqno(3.6)$$ and
$$\phi = 0 \qquad {\rm as} \quad \xi \to \infty \, . \eqno(3.7)$$
The parameter $\delta$ can vary between 0 and 2.  In the
limit $\delta \to 2$, the two vacuum states become degenerate
(this limit is where the traditional thin wall limit is
applicable).  In the opposite limit $\delta \to 0$, the
height of the barrier between the two vacua vanishes.
A typical potential for an intermediate value of $\delta$
is shown in Figure 1.

\bigskip
\centerline{\bf A. Quantum Tunneling at Zero Temperature}
\medskip

In this subsection, we evaluate the Euclidean action for the
$O(4)$ symmetric case, which determines the quantum tunneling
rate at zero temperature.  Using the rescaling of the problem
described above, we can write the $O(4)$ symmetric Euclidean
action as
$$\seA = {\pi^2 \over 2 \lambda} \int _0^\infty \, \xi^3 \, d\xi \,
\Bigl[ {1 \over 2} \Bigl( {d \phi \over d\xi} \Bigr)^2 +
{\widetilde V} [\phi(\xi)] \Bigr] \, \equiv {\pi^2 \over 2 \lambda}
{\cal B}_4(\delta) \, , \eqno(3.8)$$
where in the second equality we have defined a reduced
action ${\cal B}_4(\delta)$ which depends only on the parameter $\delta$.
We have also defined a non-dimensional (reduced) potential
$$\widetilde V \equiv {1 \over 4} \phi^4 - \phi^3 + {\delta \over 2}
\phi^2 .  \eqno(3.9)$$
As mentioned earlier, the relevant range of the parameter $\delta$ is
$$0 \le \delta \le 2 , \eqno(3.10)$$
where $\delta = 0$ corresponds to a vanishing potential barrier
height and $\delta = 2$ corresponds to the limit of degenerate
vacuum states.  For each value of $\delta$ in this range, there
exists a single solution $\phi_\delta (\xi)$ to the differential
equation (3.4) for the boundary conditions (3.6 -- 3.7).
This solution can be used to calculate the corresponding
reduced action ${\cal B}_4(\delta)$.

In the thin wall limit (i.e., the limit in which the two
vacua are nearly degenerate), the solution can be obtained
analytically [9].  In this limit, the reduced action
${\cal B}_{TW4} (\delta)$ takes the relatively simple form
$${\cal B}_{TW4} (\delta) = {2 \over 3} (2 - \delta)^{-3} \,
\eqno(3.11)$$
(this form is derived in the Appendix).
We will find it convenient to express our results
in terms of the ratio $\ratA$ defined by
$$\ratA (\delta) \equiv {{\cal B}_4(\delta) \over {\cal B}_{TW4} (\delta) } =
{3 \over 2} \, (2 - \delta^2)^3 \, {\cal B}_4(\delta) \, . \eqno(3.12)$$
Although the reduced action ${\cal B}_4 (\delta)$ varies from 0 to
$\infty$ over the allowed range of the parameter $\delta$, the
ratio $\ratA$ is a relatively slowly varying function.
The thin wall approximation becomes valid as $\delta$ approaches
2, so the ratio $\ratA$ approaches unity.  In the opposite limit
in which the barrier disappears, $\delta$ approaches zero and
${\cal B}_4(\delta)$ (and hence $\ratA$) also approaches zero.

For each value of the parameter $\delta$ in the allowed range
$0 \le \delta \le 2$, we find the solution to the equation of
motion (3.4) using numerical techniques; we then use these
solutions to determine the reduced action ${\cal B}_4 (\delta)$ and
the corresponding ratio $\ratA (\delta)$.
The resulting function $\ratA (\delta)$ is shown in Figure 2.
Since $\ratA$ has relatively simple behavior, we can approximate
the function with a cubic polynomial.  In the spirit of Ref. [22],
we can write
$$\ratA (\delta) = \alpha_1 \delta + \alpha_2 \delta^2 +
\alpha_3 \delta^3 \, , \eqno(3.13)$$
where the $\alpha_j$ are constants. We find that
$\alpha_1$ = 13.832, $\alpha_2$ = --10.819, and
$\alpha_3$ = 2.0765.
These values of the constants produce an approximation
to the true (numerically determined) function with an
absolute error which is bounded
to be less than 0.004; the cubic polynomial gives the exact
values at the endpoints $\delta = 0$ and $\delta = 1$.
The errors result from using a simple
cubic polynomial to approximate the function $\ratA (\delta)$;
by using polynomials of higher order to fit the numerical
results, one can obtain even smaller bounds on the errors
at the expense of using a more complicated formula.
The estimated numerical errors in our procedure are
approximately an order of magnitude smaller than the
error due to the use of a cubic polynomial.

Putting all of the results of this section together, we
obtain the desired expression for the Euclidean action
$$\seA = {\pi^2 \over 3 \lambda} \, (2 - \delta)^{-3} \,
\Bigl\{ \alpha_1 \delta + \alpha_2 \delta^2 + \alpha_3 \delta^3
\Bigr\} \, , \eqno(3.14)$$
where the parameter $\delta = 8 \lambda b/a^2$ is determined by
the parameters in the original potential and where the constants
$\alpha_j$ are given above.  This expression is valid over the
entire range of quartic potentials, from the thin wall (nearly
degenerate) limit to the opposite limit of vanishing
barrier height.

In order to show what the instanton solutions actually look
like, we plot a sequence of solutions $\phi(\xi)$ in Figure 3
for various values of the parameter $\delta$.
As $\delta$ increases, the solutions extend to
larger and larger (non-dimensional) radii $\xi$.  As shown
in Figure 3, the solutions for $\delta < 1.5$ have no
well-defined radius, i.e., the solutions $\phi (\xi)$
vary smoothly from $\phi_0$ to the false vacuum value
$\phi = 0$.  As the value of $\delta$ approaches that of
the thin-wall limit ($\delta=2$), however, the region over
which the solution varies from $\phi_0$ to 0 becomes small
compared to the extent (in $\xi$) of the solution; hence,
the thin-wall limit becomes a reasonable description in
this case.

\bigskip
\centerline{\bf B. Thermal Tunneling at Finite Temperature}
\medskip

In this subsection, we calculate the Euclidean action in the
$O(3)$ symmetric case.  As in the previous section, we derive
a semi-analytic expression for the action which can be used to
determine the tunneling rate (see Eq. [2.4]).  The action
$\seB$ for the $O(3)$ symmetric case can be written
$$\seB = {\pi a \over 2 \lambda^{3/2} }
\int _0^\infty \, \xi^2 \, d\xi \,
\Bigl[ {1 \over 2} \Bigl( {d \phi \over d\xi} \Bigr)^2 +
{\widetilde V} [\phi(\xi)] \Bigr] \, \equiv {\pi a \over 2 \lambda^{3/2} }
{\cal B}_3 (\delta) \, , \eqno(3.15)$$
where we have used the same scaling transformation as before (Eq. [3.3]).
In the second equality, we have defined the reduced action
${\cal B}_3 (\delta)$ for the three dimensional case.  Keep in mind,
however, that the parameters $a$, $b$, and $\lambda$ appearing
in the potential are generally not the same as those in the
potential for the zero temperature limit considered previously.

In the thin wall limit (i.e., the limit of nearly degenerate
vacua), the reduced action ${\cal B}_3 (\delta)$ takes the form
$${\cal B}_{TW3} (\delta) \equiv {16 \sqrt{2} \over 81} \,
(2 - \delta)^{-2} \,  \eqno(3.16)$$
(see the Appendix).  As before, we find it convenient to
express the result in terms of the ratio $\ratB$
$$\ratB (\delta) \equiv
{{\cal B}_3 (\delta) \over {\cal B}_{TW3} (\delta) } =
{81 \over 16 \sqrt{2}} \, (2 - \delta)^2 \,
{\cal B}_3 (\delta) \, . \eqno(3.17)$$
The resulting function $\ratB (\delta)$ is shown in Figure 4.
Notice that the function $\ratB$ is relatively slowly
varying over the allowed range of the parameter $\delta$.
We find that a reasonable fit to the numerically determined
function can be found using a function of the form
$$\ratB (\delta) = \sqrt{\delta/2} \,
\Bigl\{ \beta_1 \delta + \beta_2 \delta^2 +
\beta_3 \delta^3 \Bigr\} \, , \eqno(3.18)$$
where $\beta_1$ = 8.2938, $\beta_2$ = --5.5330, and
$\beta_3$ = 0.8180. The absolute errors due to this
polynomial-type approximation are bounded
to be less than 0.033. Putting the results of this section
together, we obtain the desired semi-analytic expression for
the three dimensional Euclidean action:
$$\seB = {\pi a \over \lambda^{3/2} } \, {8 \sqrt{2} \over 81}
(2 - \delta)^{-2}  \, \sqrt{\delta/2} \,
\Bigl\{ \beta_1 \delta + \beta_2 \delta^2 +
\beta_3 \delta^3 \Bigr\}  \, , \eqno(3.19)$$
where $\delta = 8 \lambda b/a^2$ and
where the constants $\beta_j$ are given above.

\medskip
\bigskip
\goodbreak
\centerline{\bf IV. SOLUTIONS FOR QUARTIC/LOGARITHMIC POTENTIALS}
\medskip

In this section we consider the tunneling problem for a potential of
the general form
$$V ( \varphi ) = (2 A - B) \sigma^2 \varphi^2 -
A \varphi^4 + B \varphi^4
\ln \left( {\varphi^2 \over \sigma^2} \right) \, . \eqno(4.1)$$
This form of the potential arises in simplified versions of the
theory of electroweak interactions.  For example, in the one-loop
approximation, the effective potential for the Higgs boson in
the electroweak model takes this general form [23].
In this paper, however, we decouple our discussion of tunneling
from any specific model of particle physics.  We simply take
equation (4.1) as an expression for an effective potential
and then determine semi-analytic expressions for the
Euclidean action in terms of the parameters appearing in
the potential.

Since this effective potential contains only three parameters,
we can reduce the problem to a one parameter family of differential
equations by using scaling transformations analogous to those of
equation (3.3) in the previous section.  In particular, we take
$$\phi = \varphi/\sigma \qquad {\rm and} \qquad
\xi = 2 \sigma \sqrt{B} \, r \ . \eqno(4.2)$$
We thus obtain the differential equation
$${d^2 \phi \over d \xi^2} + {N - 1 \over \xi} {d \phi \over d \xi}
= \delta (\phi - \phi^3) + \phi^3 \ln \phi^2 \, , \eqno(4.3)$$
where $N$ = 3, 4 as before and where the remaining parameter $\delta$ is
defined in terms of the original parameters in the potential through
$$\delta \equiv {2 A - B \over 2 B} \ .  \eqno(4.4)$$
The boundary conditions are given by equations (3.6) and (3.7).
Proceeding as in \S III, we define a reduced action
${\cal B}_N (\delta)$ through
$${\cal B}_N (\delta ) = \int^\infty_0 \xi^{N-1} d \xi
\left[ {1 \over 2} \bigl( {d \phi \over d \xi } \bigr)^2 +
\widetilde{V} \bigl[ \phi(\xi) \bigr] \right] \ ,
\eqno(4.5)$$
and hence
$$S_{E4} = {\pi^2 \over 2B} {\cal B}_4 (\delta ) \qquad
{\rm and} \qquad S_{E3} = {2 \pi \sigma \over \sqrt{B}}
{\cal B}_3 (\delta ) \, . \eqno(4.6)$$
The reduced potential $\widetilde{V} (\phi )$ can be written
$$\widetilde{V} (\phi) = {1 \over 4} \left\{
\phi^4 \ln \phi^2 + 2 \delta \phi^2 - \left(\delta + {1 \over 2}\right)
\phi^4 \right\} \, , \eqno(4.7)$$
and has minima at $\phi = 0$ and at $\phi = 1$.  The relevant
range of the parameter $\delta$ is
$$0 \leq \delta \leq {1 \over 2} \, , \eqno(4.8) $$
where the limit $\delta = 0$ implies no potential barrier and the
opposite limit $\delta = {1 / 2}$ implies degenerate minima.  In
this latter (degenerate) limit, the thin-wall approximation is valid
and we obtain the closed-form solution for the reduced action
${\cal B}_{TW4}$ for the $N = 4$ case (see the Appendix),
$${\cal B}_{TW4} =  54 I^4 (1-2 \delta)^{-3} \ . \eqno(4.9)$$
Here, $I$ is a pure number and is defined by the integral
$$I \equiv \int^1_0 dx \, [1 - x + x \ln x]^{1/2} \approx 0.419900
\ , \eqno(4.10)$$
where the numerical estimate has been obtained computationally.
Similarly, we find the reduced action ${\cal B}_{TW3} (\delta )$
for the $N = 3$ case (see the Appendix),
$${\cal B}_{TW3} (\delta) = {8 \over 3} \sqrt{2} \, I^3 \,
(1-2 \delta)^{-2} \, . \eqno(4.11)$$
Next we define the ratio of the reduced action to
that of the thin wall limit:
$${\cal R}_4 (\delta) \equiv
{{\cal B}_4 (\delta) \over {\cal B}_{4TW} (\delta )} =
{(1-2 \delta)^3 \over 54 I^4} \ {\cal B}_4 (\delta) \, , \eqno(4.12)$$
$${\cal R}_3 (\delta) \equiv
{{\cal B}_3 (\delta) \over {\cal B}_{3TW} (\delta )} =
{3 (1-2 \delta)^2 \over 8 \sqrt{2} I^3} \ {\cal B}_3 (\delta)
\, . \eqno(4.13)$$

The functions ${\cal R}_N (\delta )$ have been calculated numerically for
the potential of equation (4.1).  The results are shown in Figure 5 for
${\cal R}_4 (\delta )$ and in Figure 6 for ${\cal R}_3 (\delta )$.
As in the case of purely quartic potentials (\S III), the functions
${\cal R}_N (\delta)$ are relatively slowly varying over
the allowed range of the parameter $\delta$.  For the $N=4$ case,
the function ${\cal R}_4$ can be fit with a function of the form
$${\cal R}_4 (\delta) = (2 \delta)^{n_\gamma}
\Bigl\{ 1 + \gamma_1 \delta + \gamma_2 \delta^2 +
\gamma_3 \delta^3 \Bigr\} \, , \eqno(4.14)$$
where $n_\gamma$ = 0.30, $\gamma_1$ = 2.0151, $\gamma_2$ = --5.5915,
and $\gamma_3$ = 3.1225.  This expression produces an approximation
to the true (numerically determined) function with errors bounded
to be less than 0.0095.  For the $N=3$ case,
the function ${\cal R}_3$ can be fit with a function of the form
$${\cal R}_3 (\delta) = (2 \delta)^{n_\mu}
\Bigl\{ 1 + \mu_1 \delta + \mu_2 \delta^2 +
\mu_3 \delta^3 \Bigr\} \, , \eqno(4.15) $$
where $n_\mu$ = 0.557, $\mu_1$ = 4.2719, $\mu_2$ = --14.5908,
and $\mu_3$ = 12.0940.  This expression produces an approximation
to the true (numerically determined) function with errors bounded
to be less than 0.0130.

Putting all of the results of this section together, we
obtain the desired semi-analytic expressions for the
Euclidean action in both four and three dimensions:
$$S_{E4} = {27 \pi^2 I^4 \over B \, (1 - 2 \delta)^3 }
\, (2 \delta)^{n_\gamma} \Bigl\{ 1 + \gamma_1 \delta + \gamma_2
\delta^2 + \gamma_3 \delta^3 \Bigr\} \, , \eqno(4.16)$$
$$S_{E3} = {16 \pi \sigma I^3 \over 3 \, (1 - 2 \delta)^2 }
\Big( {2 \over B } \Bigr)^{1/2} \, (2 \delta)^{n_\mu}
\Bigl\{ 1 + \mu_1 \delta + \mu_2 \delta^2 +
\mu_3 \delta^3 \Bigr\} \, . \eqno(4.17)$$
The parameter $\delta = A/B - 1/2$ and the constants $I$, $n$,
$\gamma_j$, and $\mu_j$ are given above. These expressions are
valid for the range $0 \le \delta \le 1/2$.

\bigskip
\centerline{\bf V. DISCUSSION}
\medskip

In this paper, we have derived semi-analytic expressions for the
Euclidean action for the theory of a single scalar field with a
quartic potential.  We have derived similar expressions for
effective potentials with quartic/logarithmic terms.  For
these potentials, we have found the action for both
the four-dimensional $O(4)$ symmetric case (Eqs. [3.14] and [4.16])
and for the three-dimensional $O(3)$ symmetric case (Eqs. [3.19]
and [4.17]). These results can be used to determine the decay probability
per unit time per unit volume for both quantum (zero temperature)
tunneling (see Eq. [2.1]) and for tunneling at finite temperature
(see Eq. [2.7]).  Our expressions are valid over the entire
range of potentials, from the limit of nearly degenerate vacua
(where the thin-wall limit is applicable) to the opposite limit
of vanishing potential barrier height [24].

Although many workers in the field have used the expression
for the Euclidean action in the thin wall limit, the results
of this paper show that the thin wall limit does not provide
a good approximation unless the potential is very nearly
degenerate.  As shown in both Figures 2 and 4 for the case
of quartic potentials, the ratio ${\cal R}_N$ grows rather
rapidly as $\delta$ is decreased below the value corresponding
to degenerate vacua $(\delta=2)$.   For the effective potential
considered in \S IV, the ratio ${\cal R}_N$ remains close to
unity over a larger fraction of the allowed range of the
parameter $\delta$ (see Figures 5 and 6) and thus the thin-wall
limit provides a somewhat better approximation in this case.

The departure of the Euclidean action from its value in the
thin wall limit can be important in applications.  Consider,
for example, the case of the inflationary universe.  For
inflationary models which proceed through a first order
phase transition, the quantity that determines the dynamics
is the bubble nucleation efficiency $\eff$, defined by
$$\eff \equiv {K \over H^4} \, {\rm e}^{-\seA} \approx
{1 \over 64} \, \Bigl[ {m_{\rm Pl} \over \eta} \Bigr]^4 \,
{\rm e}^{-\seA} \, , \eqno(5.1)$$
where $H$ is the Hubble parameter (see, e.g., Refs. [1--6]).
In the second approximate equality, $\eta$ is the energy
scale of the inflationary phase transition (often the GUT
scale) and $m_{\rm Pl}$ is the Planck scale.  In order for
successful inflation to take place, the nucleation efficiency
$\eff$ must start with a small value $\sim$ $10^{-4}$ and then
increase to be of order unity at the end of the inflationary
epoch [1--6].  For  $m_{\rm Pl}$/$\eta$ = $10^3$, we thus
require
$$10^{-10} \le {\rm e}^{-\seA} \le 10^{-6} \, , \eqno(5.2)$$
which clearly shows the need for an accurate determination
of the action $\seA$; a factor of two error in the action can
produce a large effect when ${\rm e}^{-\seA}$ is small.

Although the results of this paper are only valid for the
particular forms of the potentials considered here,  we note
that other (non-quartic) potentials with two minima can often
be reasonably well fit with quartic polynomials. The results
of this paper can be used in such cases by approximating the
true potential with a quartic form, although the degree
of accuracy will be somewhat reduced. We also note that
the results of this paper are limited to the case of $\lambda > 0$
(i.e., positive quartic terms in the potential).  The well known
solution [25] for the potential $V = - \lambda \varphi^4$ is not
covered here, but the procedure of this paper can be easily
adapted to find the action for potentials with negative
quartic terms. Finally, we note that the procedure of this paper
can be generalized even further; by rescaling the field and the
radial coordinate (as in Eqs. [3.3]), we can eliminate two
parameters from any polynomial potential [26].

\vskip 0.50truein
\centerline{\bf Acknowledgements}
\vskip 0.10truein

I would like to thank A. Dolgov, K. Freese, G. Kane,
I. Rothstein, and R. Watkins for useful discussions.
This work was supported by NASA Grant No. NAGW--2802 and by
funds from the Physics Department at the University of Michigan.

\vskip 0.50truein
\centerline{\bf APPENDIX: THE THIN WALL LIMIT}
\medskip

In this Appendix, we review the calculation of the Euclidean action in
the limit that the two vacuum states are nearly degenerate.  In this
limit (generally known as the ``thin wall limit''), the calculation
can often be done analytically [9, 10, 11].  We begin by conceptually
dividing the reduced potential ${\widetilde V} (\phi )$ into two parts
$${\widetilde V} (\phi) = V_0 (\phi ) + V_\epsilon (\phi )
\, , \eqno ({\rm A}1)$$
where $V_0$ is the form of the potential when the vacua are exactly
degenerate and $V_\epsilon$ is the part of the potential that
provides the asymmetry between vacua (this term is assumed to be small
in this limit).  The reduced action ${\cal B}_N$ (the quantity we want
to approximate), can be written
$${\cal B}_N = \int^\infty_0 \xi^{N-1} d \xi \,
\Bigl[ {1 \over 2} \bigl( {d \phi \over d \xi} \bigr)^2 +
\widetilde{V} [\phi (\xi)] \Bigl] \ , \eqno({\rm A}2)$$
where $N = 3$ or 4 and where $\widetilde{V} (\phi)$ is the reduced
potential as defined in the text.

We now assume that the radius $R$
of the bubble is large compared to its thickness.  In terms of the
function $\phi (\xi)$, this approximation assumes that the interval
$\Delta \xi$ over which the field $\phi$ changes from its value in one
vacuum state to its value in the other vacuum state is small compared
to  the value $\xi \cong R$ where this change occurs.  We therefor divide
the integral for ${\cal B}_N$ into 3 parts: $\xi < R$, $\xi \sim R$, and
$\xi > R$.  In the first region $\xi < R$, the field $\phi$ is sitting
in the true vacuum state and
$${\widetilde V} (\phi ) = - \epsilon \qquad {\rm and} \qquad
{d \phi \over d \xi} = 0 \ . \eqno({\rm A}3)$$
The contribution to the action for this region is simply
$$\int^{R}_0 \xi^{N-1} d \xi \, [{\widetilde V} (\phi_T)] =
- {R^N \over N} \epsilon \ . \eqno({\rm A}4)$$
In the second region $\xi \sim R$, we assume that the field $\phi$
changes its value from one vacuum state to the other over an
interval $\Delta \xi$ that is small compared to the radius $R$.
With this assumption, the contribution to the action for this
region becomes
$$\int_{\Delta \xi} \xi^{N-1} d \xi \,
\Bigl[ {1 \over 2} \bigl( {d \phi \over d \xi} \bigr)^2 +
\widetilde{V} [\phi (\xi)] \Bigl] \, = \, R^{N-1} S_1 \, ,
\eqno({\rm A5a})$$
where we have defined the dimensionless one-dimensional action
$S_1$ in the second equality.  We also assume that $R$ is large enough
that the ``friction'' term $\sim {d \phi / d \xi}$ in the equation
of motion for $\phi$ can be neglected (see Eq. [3.4]).
In this case, we obtain the expression
$$S_1 = \int d \xi \, 2 V_0 (\phi ) =
\int_{\phi_F}^{\phi_T} d \phi \sqrt{2 V_0} \, , \eqno({\rm A5b})$$
where the integral in $\phi$ extends from the false vacuum
state to the true vacuum state and where we have assumed
${\widetilde V} \approx V_0$ (see Ref. [9] for further discussion).
Finally, in the region $\xi > R$, the field $\phi$ is in its false
vacuum state where $V = 0$ (and $d\phi/d\xi$ = 0).  Thus, we
get no contribution to the action in this region.

Putting the above results together, we obtain the approximation
$${\cal B}_{TWN} =  R^{N-1} S_1 - \epsilon R^N / N \ , \eqno({\rm A}6)$$
where the bubble radius $R$ remains unspecified. To determine the
value of $R$, we extremize ${\cal B}_{TWN}$ with respect to $R$ and obtain
$$R = (N - 1) S_1 / \epsilon \, . \eqno({\rm A}7)$$
We now use this expression for $R$ to obtain the reduced
action ${\cal B}_{TWN}$:
$${\cal B}_{TWN} = {(N - 1)^{N-1} \over N}
{S^N_1 \over \epsilon^{N-1}} \ . \eqno({\rm A}8)$$
Now all that remains is to find $S_1$ and $\epsilon$ for the
potentials considered in this paper.

For the quartic potential (\S III), we write the reduced
potential in the form
$$\widetilde{V} = {1 \over 4} \phi^2 (\phi - 2)^2 - {1 \over 2}
(2 - \delta ) \phi^2 \ , \eqno({\rm A}9)$$
where $\delta = 8 \lambda b/ a^2$ and varies over the range
$0 \le \delta \le 2$.  The first term in equation (A9)
represents the degenerate part ${V}_0$ of the potential.
The second term provides the asymmetry between the vacuum
states and implies that $\epsilon = 2 (2 - \delta )$.  Notice
that the false vacuum is at $\phi = 0$ and the true vacuum is
at $\phi = 2$ in the nearly degenerate limit (small $\epsilon$).
For this potential, we find that the one-dimensional action is
given by
$$S_1 = {1 \over \sqrt{2}} \int^2_0 (2 - \phi) \phi \, d \phi
= {2 \sqrt{2} \over 3} \ , \eqno({\rm A}10)$$
and hence the reduced action becomes
$${\cal B}_{TWN} = {(N-1)^{N-1} \over N \, 3^N} {2^{1+N/2} \over
(2 - \delta )^{N-1}} \ . \eqno({\rm A}11)$$
The expressions for $N = 4$ and 3 have been used in equations
(3.11) and (3.16), respectively, in the text.

For the effective potential of \S IV (see Eq. [4.1]),
we write the reduced potential as
$$\widetilde{V} = {1 \over 4} \left\{ \phi^4 \ln \phi^2 +
2 \delta \phi^2 - \left( \delta + {1 \over 2} \right)
\phi^4 \right\} \ , \eqno({\rm A}12)$$
where $\delta = A/B - 1/2$ and varies over the range
$0 \le \delta \le 1/2$. In this case, the false vacuum is at
$\phi = 0$ and the true vacuum is at $\phi = 1$ for all values
of $\delta$.  The energy difference $\epsilon$ is given by
$$\epsilon = - \widetilde{V} (\phi = 1) =
{ (1 - 2 \delta) \over 8} \ . \eqno({\rm A}13)$$
The degenerate potential ${V}_0$ is given by equation (A12) evaluated
with $\delta$ = 1/2 .  Using these results, we find that
$$S_1 = {1 \over \sqrt{2}} \int^1_0 \phi d \phi \,
[ \phi^2 \ln \phi^2 + 1 - \phi^2 ]^{1/2} =
{\sqrt{2} \over 4} I \, , \eqno({\rm A}14)$$
where we have defined the dimensionless quantity $I$ through
$$I \equiv \int^1_0 dx [1 - x + x \ln x ]^{1/2} \ .
\eqno({\rm A}15)$$
As discussed in the text, $I$ has a value of $\approx$ 0.419900.

With the above results, we find the desired expression for the
reduced action in the thin wall limit:
$${\cal B}_{TWN} = {(N - 1)^{N-1} \over 8 N}
{I^N 8^{N/2} \over (1-2 \delta)^{N-1}} \ . \eqno({\rm A}16)$$
The expressions for $N = 4$ and 3 have been used in equations (4.9)
and (4.11), respectively, in the text.

\vskip 1.0truein
\centerline{\bf REFERENCES}
\vskip 0.10truein

\item{[1]} A. H. Guth, {\sl Phys. Rev.} D {\bf 23}, 347 (1981).

\item{[2]} A. H. Guth and E. J. Weinberg, {\sl Nucl. Phys.} {\bf B}
{\bf 212}, 321 (1983).

\item{[3]} D. La and P. J. Steinhardt, {\sl Phys. Rev. Lett.}
{\bf 376}, 62 (1989); D. La and P. J. Steinhardt,
{\sl Phys. Lett.}, {\bf 220 B}, 375 (1989).

\item{[4]} F. C. Adams and K. Freese,
{\sl Phys. Rev.} D {\bf 43}, 353 (1991);
A. Linde, {\sl Phys. Lett.} {\bf B} {\bf 249}, 18 (1990).

\item{[5]} P. J. Steinhardt and F. S. Accetta,
{\sl Phys. Rev. Lett.} {\bf 64}, 2740 (1990).

\item{[6]} M. S. Turner, E. J. Weinberg, and L. M. Widrow,
{\sl Phys. Rev.} D {\bf 46}, 2384 (1992).

\item{[7]} See, e.g., the review of
A. D. Dolgov, {\sl Phys. Rep.}, in press; see also
N. Turok, {\sl Phys. Rev. Lett.} {\bf 68}, 1803 (1992);
M. Dine, R. Leigh, P. Huet, A. Linde, and D. Linde,
{\sl Phys. Rev.} D {\bf 46}, 550 (1992).

\item{[8]} A. D. Linde, {\sl Phys. Lett.} {\bf B 70}, 306 (1977);
A. D. Linde, {\sl Phys. Lett.} {\bf B 92}, 199 (1980);
A. H. Guth and E. J. Weinberg, {\sl Phys. Rev. Lett.} {\bf 45}, 1131 (1980);
P. J. Steinhardt, {\sl Nucl. Phys.} {\bf B 179}, 492 (1981);
R. A. Flores and M. Sher, {\sl Phys. Rev.} D {\bf 27}, 1679 (1983);
P. Arnold, {\sl Phys. Rev.} D {\bf 40}, 613 (1989).

\item{[9]} S. Coleman, {\sl Phys. Rev.} D {\bf 15}, 2929 (1977).

\item{[10]} M. B. Voloshin, I. Yu. Kobzarev, and L. B. Okun,
{\sl Sov. J. Nucl. Phys.}, {\bf 20}, 644 (1975).

\item{[11]} S. Coleman, {\sl Aspects of Symmetry}
(Cambridge Univ. Press, Cambridge, England, 1985).

\item{[12]} The nucleation of topological defects
during inflation has been studied by R. Basu,
A. H. Guth, and A. Vilenkin, {\sl Phys. Rev.} D {\bf 44}, 340 (1991);
R. Basu and A. Vilenkin, {\sl Phys. Rev.} D {\bf 46}, 2345 (1992).

\item{[13]} K. Lee and E. J. Weinberg,
{\sl Nucl. Phys.}, {\bf B 267}, 181 (1986).

\item{[14]} M. J. Duncan and L. G. Jensen,
{\sl Phys. Lett.} {\bf B 291}, 109 (1992).

\item{[15]} A. Linde, {\sl Nucl. Phys.} {\bf B 216}, 421 (1983).

\item{[16]} C. Callen and S. Coleman,
{\sl Phys. Rev.} D {\bf 16}, 1762  (1977).

\item{[17]} Euclidean time is related to usual time through
$t_E = i t$.  The tunneling rate is associated with the equation
of motion in imaginary time because the imaginary part of the
energy is responsible for the decay rate; see, e.g.,
Refs. [9, 11] for further discussion.

\item{[18]} S. Coleman, V. Glasser, and A. Martin,
{\sl Comm. Math. Phys.}, {\bf 58}, 211 (1978).

\item{[19]} S. Weinberg, {\sl Phys. Rev.} D {\bf 9}, 3357 (1974);
L. Dolan and R. Jackiw,  {\sl Phys. Rev.} D {\bf 9}, 3320 (1974);
D. A. Kirzhnits and A. D. Linde, {\sl Ann. Phys.} {\bf 101}, 195 (1976);
A. D. Linde, {\sl Rep. Prog. Phys.} {\bf 42}, 389 (1979).

\item{[20]} The opposite case with a positive coefficient for
the cubic term and the true vacuum at negative $\varphi$ can be
obtained by a reflection $\varphi \to - \varphi$, which does not
change the dynamics.

\item{[21]} When gravitational effects are considered, the
absolute scale of the false vacuum plays a role, but the effect
is ${\cal O} \bigl[ (\eta/m_{\rm Pl})^4 \bigr]$, where $\eta$ is the
energy scale of the potential and $m_{\rm Pl}$ is the Planck mass;
see S. Coleman and F. De Luccia, {\sl Phys. Rev.} D {\bf 21},
3305 (1980).

\item{[22]} M. Abramowitz and I. A. Stegun,
{\sl Handbook of Mathematical Functions} (Dover, New York, 1965).

\item{[23]} S. Weinberg, {\sl Phys. Rev. Lett.} {\bf 36}, 294 (1976);
A. Salam, in Elementary Particle Theory (Nobel Symp. No. 8) ed.
N. Svartholm (Almqvist and Wiksell, Stockholm, 1968) 367;
A. D. Linde, {\sl JETP Lett.} {\bf 23}, 64 (1976);
S. Coleman and E. J. Weinberg, {\sl Phys. Rev.} D {\bf 7}, 1888 (1973).

\item{[24]} Keep in mind, however, that the semi-classical techniques
used in this paper can break down when the potential barrier height
vanishes (when the parameter $\delta$ is nearly zero).

\item{[25]} S. Fubini, {\sl Nuovo Cim.} {\bf 34A}, 525 (1976).

\item{[26]} The procedure of this paper itself is a generalization
of that of Ref. [15], where potentials with only two parameters
are introduced and rescalings are used to reduce the calculation
to a single numerical problem.  Notice that in the limit that
$\lambda \to 0$ (and hence $\delta \to 0$), the expressions for
the action given in equations (3.14) and (3.19) for quartic
potentials are nonzero.  These limiting expressions are the
correct form of the Euclidean action for tunnelling in a cubic
potential.  This problem has been done in Ref. [15] and the
results are in numerical agreement with those obtained here.

\newpage
\vskip 1.0truein
\centerline{\bf FIGURE CAPTIONS}
\vskip 0.10truein

\noindent
Figure 1. A generic quartic potential $V(\phi)$.
The potential has been shifted both vertically and horizontally
to place the false vacuum at the origin ($\phi = 0$, $V = 0$).

\vskip 0.10truein
\noindent
Figure 2. The ratio $\ratA (\delta)$ of the true Euclidean
action for the $O(4)$ symmetric theory to that obtained in
the ``thin wall limit''.  This function is valid for all
quartic potentials, from the limit of nearly degenerate
vacua (at $\delta = 2$) to the opposite limit of vanishing
barrier height (at $\delta = 0$). The dashed line at
$\ratA = 1$ corresponds to the thin wall limit.

\vskip 0.10truein
\noindent
Figure 3. Instanton solutions $\phi(\xi)$ for values of
$\delta$ = 1.0, 1.5, 1.9, and 1.95. The solutions extend
to larger values of $\xi$ for larger values of $\delta$.

\vskip 0.10truein
\noindent
Figure 4. The ratio $\ratB (\delta)$ of the true Euclidean
action for the $O(3)$ symmetric theory to that obtained in
the ``thin wall limit''.  This function is valid for all
quartic potentials, from the limit of nearly degenerate
vacua (at $\delta = 2$) to the opposite limit of vanishing
barrier height (at $\delta = 0$).  The dashed line at
$\ratB = 1$ corresponds to the thin wall limit.

\vskip 0.10truein
\noindent
Figure 5. The ratio $\ratA (\delta)$ for the effective potential
of \S IV.  Here, $\ratA$ is the ratio of the true Euclidean
action for the $O(4)$ symmetric theory to that obtained in
the ``thin wall limit''.  The dashed line at $\ratA = 1$
corresponds to the thin wall limit.

\vskip 0.10truein
\noindent
Figure 6. The ratio $\ratB (\delta)$ for the effective potential
of \S IV.  Here, $\ratB$ is the ratio of the true Euclidean
action for the $O(3)$ symmetric theory to that obtained in
the ``thin wall limit''.  The dashed line at $\ratB = 1$
corresponds to the thin wall limit.

\bye